\documentstyle [12pt,epsfig]{article} 
\makeatletter
\@addtoreset{equation}{section}
\makeatother

\newcommand{\ba}{\begin{eqnarray}}
\newcommand{\ea}{\end{eqnarray}}

\begin{document}
\newcommand{\BS}{\bigskip}
\newcommand{\SECTION}[1]{\BS{\large\section{\bf #1}}}
\newcommand{\SUBSECTION}[1]{\BS{\large\subsection{\bf #1}}}
\newcommand{\SUBSUBSECTION}[1]{\BS{\large\subsubsection{\bf #1}}}

\begin{titlepage}
\begin{center}
\vspace*{2cm}
{\large \bf
TWO NOVEL SPECIAL RELATIVISTIC EFFECTS: SPACE  DILATATION AND TIME CONTRACTION}  
\vspace*{1.5cm}
\end{center}
\begin{center}
{\bf J.H.Field }
\end{center}
\begin{center}
{ 
D\'{e}partement de Physique Nucl\'{e}aire et Corpusculaire
 Universit\'{e} de Gen\`{e}ve . 24, quai Ernest-Ansermet
 CH-1211 Gen\`{e}ve 4.
}
\end{center}
\vspace*{2cm}
\begin{abstract}
 The conventional discussion of the observed distortions of space and time in
 Special Relativity (the Lorentz-Fitzgerald Contraction and Time Dilatation)
 is extended by considering observations, from a stationary frame, of : (i)
 objects moving with 
 constant velocity and uniformly illuminated during a short time $\tau_L$
 (their `Luminous Proper Time') in their rest frame; these may be
 called `Transient Luminous Objects' and (ii)
 a moving, extended, array of synchronised `equivalent clocks' in a common
 inertial frame. Application of the Lorentz Transformation to (i) shows
 that such objects, observed from the stationary frame
 with coarse time resolution in a direction perpendicular to their 
 direction of motion are seen
 to be at rest but  
 {\it longer} in the direction of the relative velocity $\vec{v}$ by 
 a factor $1/\sqrt{1-(v/c)^2}$ (Space Dilatation) and to (ii) that the
 moving equivalent clock at any fixed position in the
 rest frame of
 the stationary observer is seen to be running {\it faster} than a similar
 clock at rest by the factor $1/\sqrt{1-(v/c)^2}$ (Time Contraction).
 All four space-time `effects' of Special Relativity are simply classified
 in terms of the projective geometry of space-time, and the close analogy
 of these effects to linear spatial  perspective is pointed out.

\vspace*{1cm}
\par Published in: {\it American Journal of Physics {\bf 68} (2000), 267-274.} 
\end{abstract}
\end{titlepage}
 
\SECTION{\bf{Introduction}}
 In his 1905 paper on Special Relativity~\cite{x1}Einstein showed that
  Time Dilatation (TD) and the Lorentz-Fitzgerald Contraction (LFC), which had
  previously been introduced in a somewhat {\it ad hoc} way into Classical 
  Electrodynamics, are simple consequences of the Lorentz Transformation (LT),
  that is, of the geometry of space-time. 
  \par As an example of the LFC Einstein stated that a sphere moving with
   velocity $v$ would, `viewed from the stationary system', appear to be 
   contracted by the factor $\sqrt{1-(\frac{v}{c})^2}$ in its direction of
    motion where $c$ is the velocity of light in free space. It was only pointed
    out some 54 years later that if `viewed' was interpreted in the conventional
    sense of `as seen by the eye, or recorded on a photograph' then the sphere 
 does not at all appear to be contracted, but is still seen as a sphere with 
 the same dimensions as a stationary one and at the same position~\cite{x2,x3,x4} !
 It was shown in general~\cite{x3,x4} that transversely viewed moving objects
 subtending a small solid angle at the observer appear to be not distorted in shape
 or changed in size, but rather rotated, as compared to a similarly viewed and 
 orientated object at rest. This apparent rotation is a consequence of three distinct
 physical effects:
 \begin{itemize}
 \item[(i)] The LFC.
 \item[(ii)] Optical Aberration.   
 \item[(iii)] Different propagation times of photons emitted by different
 parts of the moving object.   
 \end{itemize}
 The effect (ii) may be interpreted as the change in direction of photons, emitted 
 by a moving source, due to the LT between the rest frames of the source and the
 stationary observer. Correcting for (ii) and (iii), the LFC can be deduced as a 
 physical effect, if not directly observed. It was also pointed out by Weinstein
 ~\cite{x5} that if a single observer is close to a moving object then, because of
 the effect of light propagation time delays, it will appear elongated if moving 
 towards the observer and contracted (to an extent greater than the LFC) if moving 
 away. Only an object moving strictly transversely to the line of sight of a 
 close observer shows the LFC.
 \par However, the LFC itself is a physical phenomenon similar in many ways to (iii)
 above.  The human eye or a photograph taken with a fast shutter record, as a sharp image,
 the photons incident on
 it during a short resolution time $\tau_R$. That is, the image corresponds to a
 projection at an almost fixed time
 in the frame S of observation. This implies that the photons constituting the
 observed image
 are emitted at different times from the different parts,{\it along the line of sight}, of
 an extended object. As shown below, the LFC is similarly defined by a fixed time
 projection in the
 frame S. The LT then requires that the photons constituting the image of a
 moving object
 are also emitted at different times, in
 the rest frame S' of the object, from the different parts {\it along
 its direction of motion}. In the following S will, in general, denote the
 reference frame of a `stationary' observer (space-time coordinates x,y,z,t)
 while S' refers to the rest frame of an object moving with uniform
 velocity $v$ in the direction of the positive x axis 
 relative to S ( space-time coordinates x',y',z',t').  
 \par The purpose of this paper is to point out that the $t =  constant$ projection
 of the LFC (see Section 2 below) and the $x' =  constant$ projection of TD (see Section 3 below)
 are not the only physically
 distinct space-time measurements possible within Special Relativity.
 In fact, as will be demonstrated below, there are two others: Space Dilatation (SD),
 the $t' =  constant$ projection and Time Contraction (TC), the $x =  constant$ 
 projection.  All four `effects' are pure 
 consequences of the LT. The additional effects of Optical Aberration and Light 
 Propagation Delays on the appearance of moving objects and synchronised clocks
 have been extensively discussed elsewhere~\cite{x6}. 
\par Although each of the four effects may be simply derived from the projective
geometry of the space time LT, the LFC and TD give rise to 
 more easily observable physical effects, so it is not surprising that they are 
 better known, For example the LFC is essential for the physical interpretation
 of the Michelson-Morley experiment, and TD is necessary to describe the observed
 lifetimes of unstable particles decaying in flight. In contrast, the two new effects
 SD and TC seem to have no similar simple observational consequences. As pointed out
 below, the most interesting effects are likely to result from SD, which is
 necessary to describe observations of, for example, a rotating extended object
 moving with a relativistic transverse velocity. It is easy to conceive a simple
 experiment involving the observation of two synchronised clocks in space, to
 test the TC effect. Although it is clearly of interest to work out in more
 detail such examples, there is no attempt to do so in the present paper,
 which is devoted to the precise definition of the four possible space-time
 projections of the LT and a discussion of their interrelations.

 \par The $t =  constant$ projection of 
 the LFC is the space-time measurement appropriate to the `moving bodies' of Einstein's
 original paper
 and to the photographic recording technique. This medium has no intrinsic time 
 resolution and relies on that provided by a rapidly moving shutter to provide
 a clear image. The LFC `works' as a well defined physical phenomenon
 because the `measuring rod' or other physical object under observation is assumed to
 be illuminated during the whole time interval required to make an
 observation, and so constitutes a continuous source of emitted or reflected 
 photons, such that some are always available in the different space ($\Delta x'$)
 and time ($\Delta t'$) intervals in S' for every position of the rod 
 corresponding to the time interval $\Delta t = \tau_R$ around the fixed time $t$ in the 
 observer's frame S during which the observation
 is made. If, however, the physical object of interest has internal motion
 (rotation, expansion or contraction) or is only illuminated, in its rest frame S', 
 during a short time interval, the above conditions, that assure that the 
 $t =  constant$ projection gives a well defined space-time 
 measurement no longer apply. All such objects, uniformly illuminated for a restricted
 time $\tau_L$ (their `Luminous Proper Time') in their rest frames,  
 may be called `Transient Luminous Objects'. For such objects it is
 natural to define a length measurement by taking the $t' =  constant$ projection
 in S'. The observation, from the stationary frame S, of such objects is discussed
 in Section 2 below. 

  \par In Section 3 time measurements other than the conventional TD 
   of Special Relativity are considered. The TD phenomenon
  refers only to a local clock, in the sense that its position in the frame S' is
  invariant (say at the spatial origin of coordinates $x'=0$). However, the time recorded by
  any synchronised clock in the same inertial frame is, by definition, identical.
  Einstein used such an array of `equivalent clocks' situated at different positions
  in the same inertial frame in his original discussion of the relativity of 
  simultanaeity~\cite{x1}. The question addressed in Section 3 is: 
  What will an observer in S see if he looks not only at a given local clock in S',
  but also at other, synchronised, equivalent clocks at different positions in S',
 in comparison
  to a standard clock at rest in his own frame? It is shown that such equivalent clocks
  may be seen to run slower than, or faster than, the TD prediction for a local clock.
  In particular they may even appear to {\it run faster than the standard clock}. 
  This is an example of the Time Contraction effect mentioned above.
\par In Section 4 the analogy between the Lorentz-Fitzgerald Contraction effect
 and linear perspective in two spatial dimensions is described.
 The final Section points out how all four space-time `effects' (observed distortions
 of space-time) in Special Relativity may be described in a unified way in terms
 of projective geometry, in close analogy with the effect of linear perspective in
 the perception of space.  
    
 \SECTION{\bf{Observation of Transient Luminous Objects in Motion:
  the Space Dilatation Effect}}
\begin{figure}[htbp]
\begin{center}\hspace*{-0.5cm}\mbox{
\epsfysize10.0cm\epsffile{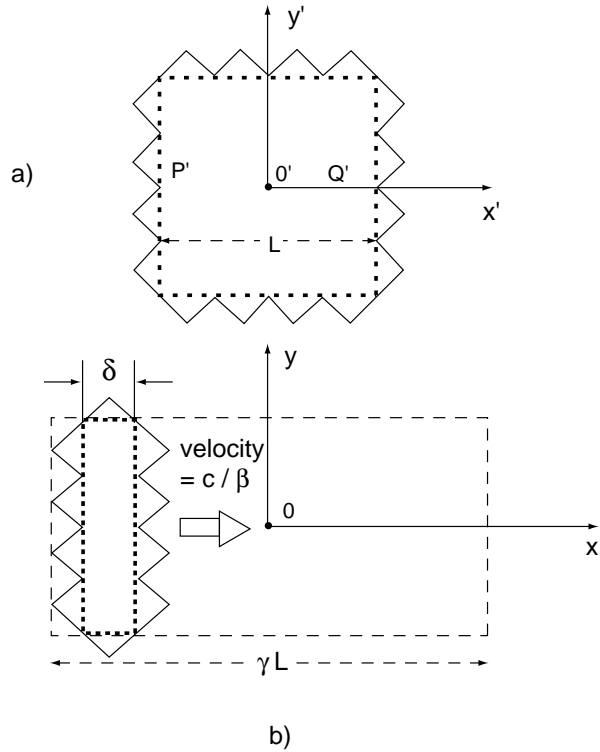}}
\caption{ a) A square `Transient Luminous Object'
 (indicated by the zig-zag outline) as viewed in the frame S' in
 which it is at rest.  b) The same object as viewed  at a fixed time
 from the frame S moving with velocity -$\beta c$ parallel to the Ox' axis.
 It is assumed that the Luminous Proper Time $\tau_L$ of the object
 is small: $\tau_L \ll \beta L/c$. The actual outlines of the
 objects are shown as short dashed lines. The long dashed rectangle of
 width $\gamma L$ in b) shows the outline of the object when
 viewed with coarse time resolution: $\tau_R \gg \gamma \beta L/c$.
 The observer in S is assumed to be sufficiently distant from the
 object that the effects of light propagation delays may be neglected.
 The object is also assumed to constitute a diffuse photon source so
 that Optical Aberration effects are negligible.}
\label{fig-fig1}
\end{center}
\end{figure}

Consider a square planar object centered at the origin of the moving frame S' as shown 
 in Fig 1a. The points P'($x'=-L/2$, $y'=0$) and Q'($x'=L/2$, $y'=0$) lie on the vertical 
edges of the square of side $L$ whose boundary is shown in Fig 1a as
the short dashed lines. Suppose now that the square is uniformly illuminated
 in the time interval $-\tau_L/2 < t' < \tau_L/2$ to give the `Transient Luminous
 Object' indicted by the zig-zag lines. The proper time interval
 $\tau_L$ is the `Luminous Proper Time' of the object. For example, the
 surface of the square may be covered with a mosaic of light-emitting diodes
 that are simultaneously switched on during a time $\tau_L$. The object as seen by an
 observer, at rest in the stationary 
 system S, viewing the object in a direction perpendicular to the plane $O'x'y'$,
 is given by the LT connecting space time points in the frame S' to 
 those in S:
\begin{eqnarray}
x & = & \gamma (x'+vt') \\
t & = & \gamma (t'+\beta \frac{x'}{c})
\end{eqnarray}
where
\[ \beta \equiv v/c,~~~~ \gamma \equiv \frac{1}{\sqrt{1-\beta^2}}. \]
 It is assumed that the stationary observer is sufficiently distant 
from the object that the effects of light propagation times are negligible, 
 and that the object is diffusely illuminated so 
that Optical Aberration effects may be neglected~\cite{x6}. In this case any
 changes in the appearance of the moving object when viewed from
 the frame S are due solely to the LT.  
 The results of the transformation for $t'=0$ and $x'=-L/2, 0, L/2$  
are given in Table 1. It can be seen that the points
 P',O',Q' are observed at different times in the frame S. This is the 
 well known effect of the relativity of simultanaeity first pointed out
 in Einstein's classic paper~\cite{x1}. It can also be seen from
 Table 1 that the distance between the positions of P' and Q' as observed
 in S is $\gamma L$; that is, the object will appear to be
 elongated if it is viewed with a time resolution larger than the difference
 in time, $\gamma \beta L/c$, between the observations in S of the space time
 points  P' and Q' that are simultaneous in the frame S'. This is the `Space
 Dilatation' (SD) effect. It will now be discussed in more detail, taking
 into account the non zero Luminous Proper Time $\tau_L$ of the Transient
 Luminous Object as well as the Resolution Time $\tau_R$ of the observer,
 so that the general conditions under which the SD effect occurs are established.
\par Space time points of the Transient Luminous Object may be observed
 at the fixed time $t$ in S provided that:
\[ x'_{MIN} < x'< x'_{MAX} \]
 where
\begin{equation}
  t = \gamma (-\frac{\tau_L}{2} + \frac{\beta x'_{MAX}}{c}) = 
\gamma (\frac{\tau_L}{2} + \frac{\beta x'_{MIN}}{c})
\end{equation}
 In (2.3) it is assumed that $x'_{MIN} > -L/2$, $x'_{MAX} < L/2$. The general condition
relating $\tau_L$,$L$, $v$ and $c$ ensuring the validity of this assumption will be
 discussed below. Using (2.1) the co-ordinates in S corresponding to $x'_{MIN}$ 
 and $x'_{MAX}$ are found to be:
\begin{eqnarray}
x_{MAX} & = & \frac{c}{\beta} (t+\frac{\tau_L}{2 \gamma}) \\
x_{MIN} & = & \frac{c}{\beta} (t-\frac{\tau_L}{2 \gamma})
\end{eqnarray}
 Thus the width $\delta$ of the Transient Luminous Object observed at time $t$ in
 S (indicated by the zig-zag lines in Fig1b; the actual boundary is shown
 by the short dashed lines) is:
\begin{equation}
\delta \equiv x_{MAX}-x_{MIN} = \frac{c \tau_L}{\beta \gamma},
\end{equation}
while, as can be seen from (2.4) and (2.5), the observer in
 S sees a luminous object that moves with 
velocity $c/\beta$, i.e. faster than than the velocity of light. 
 In the case of continous illumination of the object ($\tau_L \rightarrow \infty$)
the upper and lower limits of the object observed at the fixed time $t$ in S
 will correspond to the physical boundaries $x'_{MIN} = -L/2$, $x'_{MAX} = L/2$.
 Denoting by $t'_{MIN}$, $t'_{MAX}$ the times in S' corresponding to the 
 observation of these boundaries at time $t$ in S, then, instead of (2.3), the
 following relation is obtained:
\begin{equation}
  t = \gamma (t'_{MIN} + \frac{\beta L}{2c}) = \gamma (t'_{MAX} - \frac{\beta L}{2c})
\end{equation}
Using (2.1), the boundaries of the object observed in S at time $t$ are then: 
\begin{eqnarray}
x_{MAX} & = & \frac{L}{2 \gamma} + vt  \\
x_{MIN} & = &- \frac{L}{2 \gamma} + vt 
\end{eqnarray}
 The width of the object as seen in S is then $x_{MAX}-x_{MIN} = L/\gamma$, the
 well known LFC effect. As can be seen from (2.8) and (2.9) the object is now 
 observed to move in S with velocity $v$. Thus, in the limit $\tau_L \rightarrow \infty$
 (continous illumination of the object) the usual results of Special Relativity
 are recovered. 
\par Using (2.1) and (2.2) the upper (U) and lower (L) limits of the space time
 region in the stationary frame S swept out by the moving Transient Luminous Object
 in Fig 1b are:  
\begin{eqnarray}
x_U & = &\frac{\gamma}{2} (L+v\tau_L) \\
x_L & = &\frac{\gamma}{2} (-L-v\tau_L) \\
t_U & = &\frac{\gamma}{2} (\tau_L+\frac{\beta L}{c}) \\
t_L & = &\frac{\gamma}{2} (-\tau_L-\frac{\beta L}{c})
\end{eqnarray}
Taking account of the inequality:
\[ \frac{\beta L}{c} < \frac{ L}{v}, \]
 it can be seen that if $ \tau_L \ll \beta L/c$ the terms containing
 $\tau_L$ in (2.10)-(2.13) may be neglected, so that:
\begin{eqnarray}
x_U-x_L & \simeq & \gamma L \\  
t_U-t_L & \simeq & \frac{\gamma \beta L}{c}
\end{eqnarray}
 Thus the Space Dilatation effect of Table 1 is recovered in the limit 
 $\tau_L \rightarrow 0$.
 On the other hand, because of the inequality:
\[ v \tau_L < \frac{c \tau_L}{\beta}, \]
then, if $\tau_L \gg L/v$, the terms containing $L$  in (2.10)-(2.13) may 
be neglected, leading to the relations:
\begin{eqnarray}
x_U-x_L & \simeq & \gamma v \tau_L \\  
t_U-t_L & \simeq & \gamma \tau_L.
\end{eqnarray}
\begin{table}
\begin{center}
\begin{tabular}{|c|c c c c|} \hline
Point  & x' &  t' &  x  & t  \\ \hline
   &  &  &  &  \\
P' &  $-\frac{L}{2}$ &  0  &
 -$\frac{\gamma L}{2}$  &  -$\frac{\gamma \beta L}{2c}$ \\
   &  &  &  &  \\  
O' & 0 & 0 & 0 & 0 \\
   &  &  &  &  \\
Q' &  $\frac{L}{2}$ &  0  &
 $\frac{\gamma L}{2 }$  &  $\frac{\gamma \beta L}{2c}$ \\
  &  &  &  &  \\
\hline
\end{tabular}
\caption{ Space-time points on the object at rest in S' (see Fig. 1),
 at time $t'=0$,
 as observed in the frames S',S.}      
\end{center}
\end{table}
These are the well-known equations of Special Relativity describing
the motion of a small continously illuminated object as observed in
the frame S. The time interval $t_U-t_L$ corresponds to the TD effect and the
observed velocity is:
\begin{equation}
  (x_U-x_L)/(t_U-t_L)= \gamma v \tau_L/\gamma \tau_L =v.
\end{equation}
\par The conclusions of this Section are now summarised. When
 $ \tau_L \ll \beta L/c$ a stationary observer in S with a time resolution
 $\tau_R \ll \gamma \beta L/c$, viewing the object in 
the direction transverse to the relative velocity, sees the square object at rest in S' illuminated
 during the proper time interval $\tau_L$ as a narrow rectangular object of width
$\delta = c \tau_L/(\beta \gamma)$ moving with velocity $c/\beta$ and sweeping out
 during the time $\gamma \beta L/c$ a region of total length $\gamma L$. 
 If however the resolution time $\tau_R$ of the observer is much larger than
 $\gamma \beta L/c$ {\it the object will appear at rest but elongated by the factor
 $\gamma$ in the direction of motion}. This is the Space Dilatation effect.
  In the contrary case that the Luminous Proper Time $\tau_L$ is large ($\tau_L \gg L/v$),
 the object observed from S moves with velocity $v$  and has an apparent length $L/\gamma$
 due to the well known LFC effect. Also, in this case, the elapsed times in S and S'
 are related by the TD effect (Eqn. 2.17).
\par It should be noted that the `narrow rectangular object' referred to above
 corresponds to the case of uniform illumination of the square object. Actually,
 because of the relativity of simultaneity, different parts of the square are 
 seen at different times and positions by the stationary observer. If the square
 were illuminated using different colours: red, yellow, green, blue in four
 equal bands parallel to the $y'$ axis, in the direction of increasing $x'$, then
 the moving object in Fig 1b  would appear red during the time interval 
 $-\gamma \beta L/2c < t < \gamma \beta L/4c$, yellow during the time
 $-\gamma \beta L/4c < t  < 0$, and so on. The colours will, of course, be
 seen shifted in frequency according to the relativistic transverse Doppler effet.  

\par If the square is rotated about the $y'$ axis by an angle $\alpha$, a subtle
 interplay occurs between the effects of the LT and light propagation time delays. 
 Depending on the values of $v$ and $\alpha$ the rectangular object may
 be seen, by an observer at rest in the frame S, to  move parallel to $\vec{v}$
 (as in the case $\alpha = 0$ described above), antiparallel to $\vec{v}$, or may even
 even be stationary and of length  $\gamma L \cos \alpha$. In all
 cases the total length swept out by the object in the direction of motion
 is $\gamma L \cos \alpha$. These effects have been described in detail elsewhere~\cite{x6}.

\SECTION{\bf{Observation of an Array of Equivalent Moving Clocks: The Time Contraction Effect}}

\begin{figure}[htbp]
\begin{center}\hspace*{-0.5cm}\mbox{
\epsfysize10.0cm\epsffile{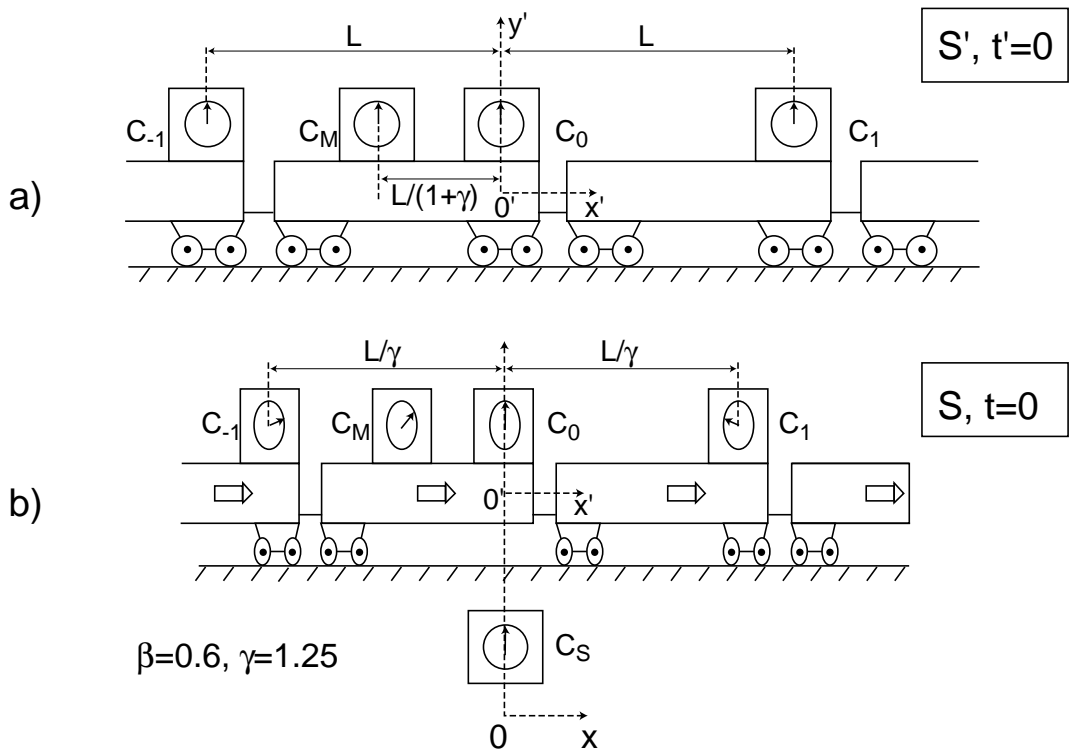}}
\caption{ a) Positions and times of equivalent clocks on the wagons of
 a train as seen by a distant observer 
 in the rest frame S' of the train at time $t' = 0$.
b) The postions and times of the same clocks as seen by a distant observer in S
 at time $t=t'=0$. The same remarks concerning light propagation time and
 Optical Aberration effects as made in the caption of Fig.1 apply.}   
\label{fig-fig2}
\end{center}
 \end{figure}
\begin{figure}[htbp]
\begin{center}\hspace*{-4.5cm}\mbox{
\epsfysize15.0cm\epsffile{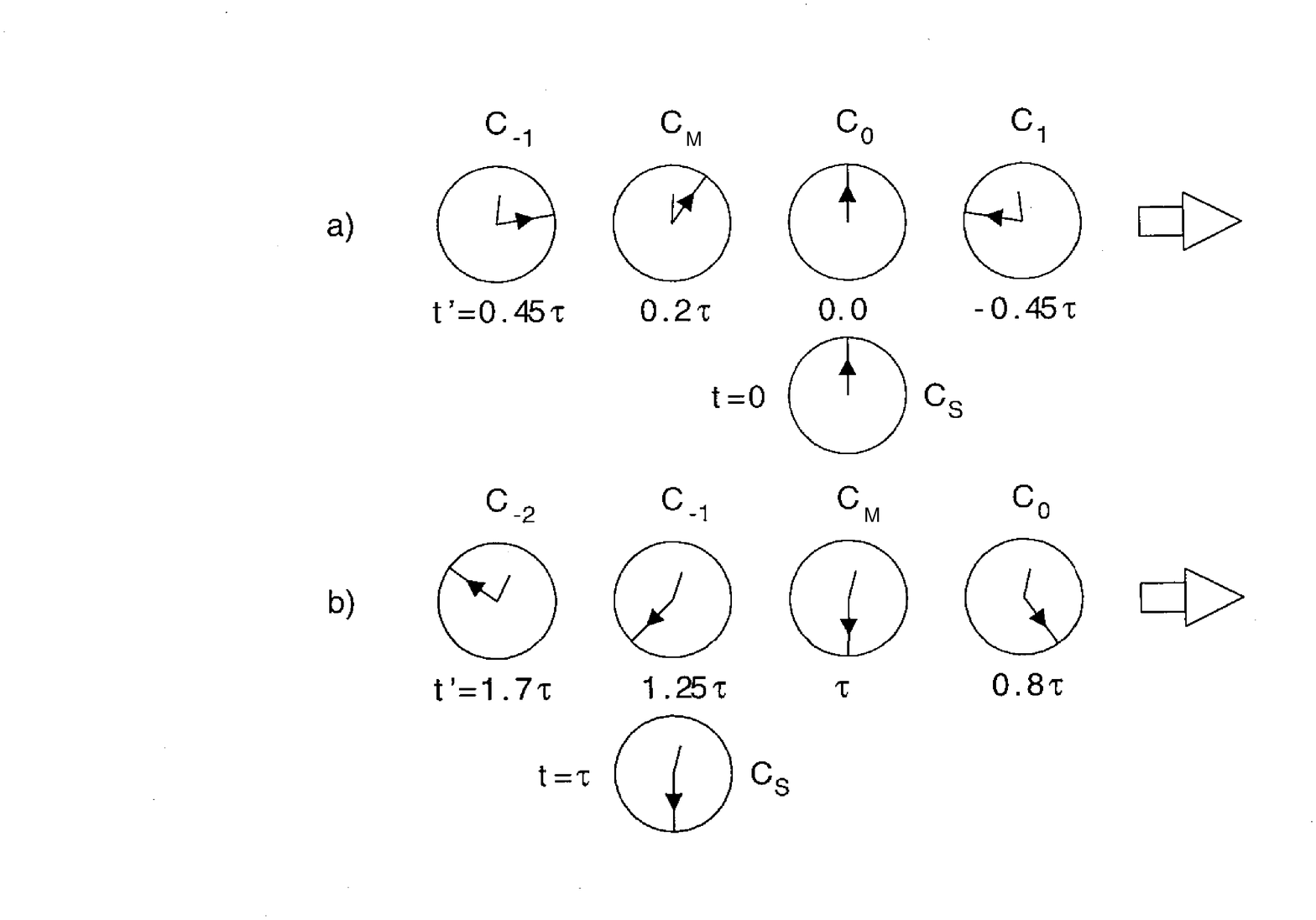}}
\caption{a) Times of equivalent clocks on the  train
 ($C_{-1}$, $C_M$, $C_0$, $C_1$) and the stationary clock $C_S$, as seen by a distant observer
 in S at time $t=0$. b) The same, at time $t=\tau$. It is assumed that $\beta = 0.6$,
 $\gamma = 1.25$.} 
\label{fig-fig3}
\end{center}
 \end{figure}

In this Section space time measurements of an array of synchronised clocks situated in the 
inertial frame S' will be considered. These clocks may be synchronised by any convenient
procedure~\cite{x7} 
 (see for example Ref.[1]). For an observer in S' all such clocks are `equivalent' in the sense
 that each of them records, independently of its position, the proper time $\tau'$ of the frame
 S'. For convenience, the array of clocks is assumed to be placed on the wagons of a
 train which is at rest in S', as shown in Fig.2a. The clocks are labelled $C_m~, m=...-2,-1,
 0,1,2,...$ and are situated (with the exception of the `magic clock' $C_M$,
 see below) at fixed distances $L$ from each other, along the Ox' axis, which is parallel 
 to the train. It is assumed that the observers in the frames S and S' view the train
 transversely at a sufficiently large distance that the effects of light propagation time
 delays may be neglected. It is clear that by considering the limit $L \rightarrow 0$
 an Equivalent Clock may be associated with each position on the train and, by extending
 the `lattice' of clocks to 3 dimensions, to any spatial position in S'.  

 The observer in S' will note that each Equivalent Clock (EC) indicates the same time,
 as shown in Fig.2a. It is now asked how the array of EC will appear to an observer at a fixed
 position in the frame S when the train is moving with velocity $\beta c$ parallel to the
 direction Ox in S (Fig.2b). It is assumed that the EC $C_0$ is placed at $x'=0$ and 
 that it is synchronised with the Standard Clock $C_S$, placed at $x=0$ in S, when
 $t=t'=0$. All the clocks are similar, that is $C_S$ and each $C_m$ record exactly
 equal time intervals when they are 
 situated in the same inertial frame.

 \par The appearence of the moving array of EC to an observer in S at $t=0$ is shown in Fig.2b, and in
 more detail in Fig.3 for both $t=0$ and $t=\tau$. The period $\tau$ is the time between
  the passage of successive EC past $C_S$.  
  The big hand of $C_S$ in Fig.3 rotates through
  $180^{\circ}$ during the time $\tau$. Explicit expressions for the observed times are
  presented in Table 2. In Fig.2b,3 the times indicated by the clocks are shown 
  for $\beta= 0.6$. These times are readily calculated using the LT equations
  (2.1),(2.2). Consider the time indicated by $C_1$ at $t=0$. The space-time points are:
  \[~~~S'~:~(L,t')~~;~~~S~:~(x,0) \] 
  Hence, Eqns.(2.1),(2.2) give:
  \begin{eqnarray}
  x & = & \gamma (L + vt')   \\
  0 & = & \gamma (t'+\frac{\beta L}{c})
  \end{eqnarray}
  which have the solution [ $C_1(t=0)$ ]:
\begin{eqnarray}
  t' & = & -\frac{\beta L}{c}   \\
  x & = & \frac{L}{\gamma}
  \end{eqnarray} 
\begin{table}
\begin{center}
\begin{tabular}{|c|c c c c c c|} \hline
 $C_S$  & \multicolumn{1}{c|}{ $C_{-2}$}
   & \multicolumn{1}{c|}{ $C_{-1}$}
   & \multicolumn{1}{c|}{ $C_{M}$} 
   & \multicolumn{1}{c|}{ $C_{0}$}
   & \multicolumn{1}{c|}{ $C_{1}$}
   & \multicolumn{1}{c|}{ $C_{2}$} \\ \cline{1-7}
   &  &  &  &  &  &  \\   
 0 & $2 \frac{(\gamma^2-1)}{\gamma} \tau$ 
   & $ \frac{(\gamma^2-1)}{\gamma} \tau$
   & $ \frac{(\gamma -1)}{\gamma} \tau$ & 0 
   & $ -\frac{(\gamma^2-1)}{\gamma} \tau$
   & $ -2 \frac{(\gamma^2-1)}{\gamma} \tau$ \\
   &  &  &  &  &  &  \\   
$\tau$ & $ \frac{(2 \gamma^2-1)}{\gamma} \tau$ 
   & $ \gamma \tau$
   & $\tau$ 
   & $\frac{\tau}{\gamma}$
   & $ -\frac{(\gamma^2-2)}{\gamma} \tau$
   & $ -\frac{(2 \gamma^2-3)}{\gamma} \tau$ \\
   &  &  &  &  &  &  \\             
\hline
\end{tabular}
\caption[]{Times observed in S of Equivalent Clocks on the moving train in Fig.2, 
at the times $t=0$ and $t=\tau$ of the stationary standard clock $C_S$. }      
\end{center}
\end{table} 
  As shown in Fig 2b, the wagons of the train appear shorter due to the LFC
  effect (Eqn.(3.4)) and also {\it the wagons at the front end of the train
  are seen at an earlier proper time than those at the rear end}. Thus a $t=0$ snapshot
   in S corresponds, not to a fixed $t'$ in S' but one which depends on $x'$: 
   $t'=-\beta x'/c$. This is a consequence of the relativity of simultaneity of space-time
   events in S and S', as first pointed out by Einstein in Ref.[1]. Here it appears in a 
   particularly graphic and striking form. Consider now the time indicated by
   $C_{-1}$ at $t=\tau$, i.e. when $C_{-1}$ is at the origin of S. The space-time points
   are:   
  \[~~~S'~:~(-L,t')~~;~~~S~:~(0,\tau) \] 
  Hence, Eqns.(2.1),(2.2) give:
  \begin{eqnarray}
  0 & = & \gamma (-L + vt')   \\
  \tau & = & \gamma (t'-\frac{\beta L}{c})
  \end{eqnarray}
  with the solutions [ $C_{-1}(t=\tau)$ ]:
\begin{eqnarray}
  t' & = & \frac{L}{v}   \\
  \tau & = & \frac{L}{\gamma v} = \frac{t'}{\gamma}
  \end{eqnarray}    
 so that
 \begin{equation}
 t'= \gamma \tau
 \end{equation}
  The EC at the origin of S at $t=\tau$ shows a later time than $C_S$ i.e.
 it is apparently running {\it faster} than $C_S$. This is an example of Time Contraction (TC).
  {\it The Time Contraction effect is exhibited by the EC observed at any fixed position in S}.
 In fact, if the observer in S can see the EC only when they are near to $C_S$ he (or she) will
 inevitably conclude that the clocks on the train run fast, not slow as in the classical TD
 effect (see below). Suppose that the observer is sitting in a waiting room with the clock
 $C_S$ and notices the time on the train (the same as $C_S$) by looking at $C_0$ as it passes 
 the waiting room window. If he (or she) then compares $C_{-1}$ as it passes the window with $C_S$
 it will be seen to be running fast relative to the latter. In order to see the TD effect
 the observer would (as will now be shown), have to note the time shown by, for example,
 $C_0$, at time $t=\tau$ as recorded by $C_S$ in comparison with that shown
 by the {\it same clock} $C_0$  at $t=0$.
 Using Eqn.(3.8),Eqn.(3.3) may be written as [$C_1(t=0)$]:
 \begin{equation}
 t' = -\beta^2 \gamma \tau = -\frac{(\gamma^2-1)\tau}{\gamma}
 \end{equation}
 This is the formula for the observed time reported in Table 2.
 Now consider $C_0$ at time $t=\tau$. The space-time points are:
 \[~~~S'~:~(0,t')~~;~~~S~:~(x,\tau) \] 
  Hence, Eqns.(2.1),(2.2) give:
  \begin{eqnarray}
  x & = & \gamma v t'   \\
  \tau & = & \gamma t'
  \end{eqnarray}
  with the solutions [ $C_0(t=\tau)$ ]:
\begin{eqnarray}
  t' & = & \tau/ \gamma \\
  x & = & v \tau  = L/\gamma
  \end{eqnarray}
 So {\it the EC $C_0$ at time $t=\tau$ indicates an earlier time, and so is 
 apparently running slower than $C_S$}.
 This is the classical Time Dilatation (TD) effect. It applies to observations
 of all {\it local clocks in S'},(i.e. those situated at a fixed value of $x'$) as well as any 
 other EC that has the same value of $x'$.
 \par As a last example consider the `Magic Clock' $C_M$ shown in Fig 2a at time $t=\tau$.
 With the space-time points:
 \[~~~S'~:~(-L/(1+\gamma),t')~~;~~~S~:~(x,\tau) \] 
  Eqns.(2.1),(2.2) give:
  \begin{eqnarray}
  x & = & \gamma [-L/(1+\gamma)+v t']   \\
  \tau & = & \gamma [ t'-\frac{\beta}{c}L/(1+\gamma)]
  \end{eqnarray}
  with the solutions [ $C_M(t=\tau)$ ]:
\begin{eqnarray}
  t' & = & \tau  \\
  x & = & \gamma v \tau / (1+\gamma) 
  \end{eqnarray}
  where the relation $L=\gamma v \tau$ from Eqn.(3.8) has been used.
  {\it Thus $C_M$ shows the same time as $C_S$ at $t=\tau$}. 
  Similar moving `Magic Clocks' can be defined that show the same time as $C_S$ at any chosen time 
  $t$ in S. Such a clock is, in general, situated at $x' = -ct(\gamma-1)/\beta \gamma$.
  All of the other clock times presented in Table 2 and shown in Figs. 2b, 3 are calculated
  in a similar way to the above examples by choosing appropriate values of $x'$ and t.
  \par The combined effects of the LT and light propagation delays for light signals
   moving parallel to the train (corresponding to observations of the array of
   Equivalent Clocks by observers on, 
    or close to the train) have been described in detail elsewhere~\cite{x6}. The
   observed spatial distortions of the train in this situation were previously 
   considered by Weinstein~\cite{x5}.  
\SECTION{\bf{Analogy with Linear Perspective in Two Dimensional Space}}
\begin{figure}[htbp]
\begin{center}\hspace*{-0.5cm}\mbox{
\epsfysize10.0cm\epsffile{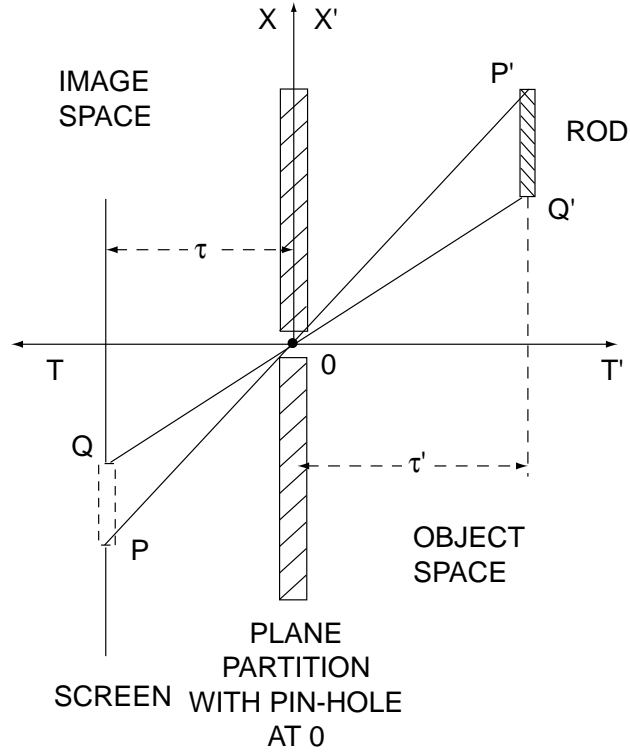}}
\caption{ An example of linear spatial perspective analagous to the
 Lorentz-Fitzgerald Contraction effect.}
\label{fig-fig4}
\end{center}
 \end{figure} 
 
\par The analogy between the observed distortions of space-time in Special Relativity
and linear spatial perspective is illustrated in Fig.4. The `Object Space' on the right 
is separated from the `Image Space' on the left by a plane partition containing a 
small aperture (pin hole). Light reflected from the rod PQ in the Object Space, can pass
 through the pin hole and produce an image on a screen located in the Image Space. 
To facilitate the comparison with the Lorentz Transformation the cartesian axes in
the Object [Image] space are denoted by (X',T') [ (X,T)] respectively (see Fig. 4). 
The T, T' axes are perpendicular to the plane of the partition and pass through the
 pin hole. The Object Space is now compared to the rest frame S' of the moving
 object, with the correspondences:
 \[ X' \Leftrightarrow x',~~~~ T' \Leftrightarrow t',  \]
 while the Image Space is compared to the frame S of the stationary observer
 with the correspondences:
 \[ X \Leftrightarrow x,~~~~ T \Leftrightarrow t,  \]
 An arbitary point with coordinates (X',T') on the rod will project into the
 Image Space the line: 
\begin{equation}
X = -\frac{X'T}{\tau'}
\end{equation}
which may be compared to the LT equation:
\begin{equation}
x = \frac{1}{\gamma}x'+vt
\end{equation}
Taking the $T = \tau = constant$ projection in (4.1), i.e. setting
the screen in the image space parallel to the planar surface at the distance 
$\tau$ from it, gives for the length $L_I$ of the image of the rod:
\begin{equation}
L_I = X_2-X_1 = \frac{\tau}{\tau'}(X_1'-X_2') = \frac{\tau}{\tau'}L
\end{equation} 
where the points 1,2 denote the ends of the rod or of its image.
Similarly taking the $t = constant$ projection in (4.2) gives, 
 for the apparent length $l_I$ of a rod, parallel to
 the x axis, of true length $l$:
\begin{equation}
l_I = x_1-x_2 = \frac{1}{\gamma}(x_1'-x_2') = \frac{l}{\gamma} 
\end{equation} 
corresponding to the LFC effect. The r\^{o}le of the factor
$1/\gamma$ in the LT is replaced, in the case of linear 
 perspective, by the ratio $\tau/\tau'$ that specifies the relative
 position and orientation of the object and the screen on which it
 is observed.
\SECTION{\bf{Discussion}}

The different space-time effects (observed distortions of space or time) in 
Special Relativity that have been discussed above are summarised in Table 3.
These are the well-known LFC and TD effects, Space Dilatation (SD) introduced
 in Section 2 above, and Time Contraction (TC) introduced in Section 3. Each 
 effect is an observed difference $\Delta q$ ($q=x,x',t,t'$) of two space or time 
 coordinates ($\Delta q = q_1-q_2$) and corresponds to a constant projection
 $\tilde{q} = constant$, i.e.
 $\Delta \tilde{q} = 0$ ($\tilde{q} \ne q$),
 in another of the four variables $x$, $x'$, $t$, $t'$ of the LT. 
 As shown in Table 3, the LFC, SD, TC and TD effects correspond, respectively,
 to constant $t$, $t'$, $x$ and $x'$ projections. After making this
 projection, the four LT equations give two relations among the remaining three
 variables. One of these describes the `space-time distortion' relating 
 $\Delta t'$ and $\Delta t$ or $\Delta x'$ and $\Delta x$ while the other gives
 the equation shown in the last column, (labelled `Complementary Effect') in 
 Table 3. These equations relate either $\Delta x$ to $\Delta t$ (for SD and TD)
 or $\Delta x'$ to $\Delta t'$ (for LFC and TC). It can be seen from the 
 Complementary Effect relations that the two space-time points defining the effect  
 (of space-time distortion) are space-like separated for LFC and SD and time-like
 separated for TC and TD.
 \par For example, for the LFC when $t_1=t_2=t$, the LT equations for the two
 space-time points are:
 \begin{eqnarray}
  x'_1 & = & \gamma (x_1-v t) \\
  x'_2 & = & \gamma (x_2-v t) \\
  t'_1 & = & \gamma (t-\frac{\beta x_1}{c}) \\
  t'_2 & = & \gamma (t-\frac{\beta x_2}{c}) 
  \end{eqnarray}
  Subtracting (5.1) from (5.2) and (5.3) from (5.4) gives:
 \begin{eqnarray}
 \Delta x' & = & \gamma \Delta x \\
 \Delta t' & = & -\frac{\gamma \beta}{c} \Delta x
 \end{eqnarray}
 Eqn.(5.5) describes the LFC effect, while combining Eqns.(5.5) and (5.6) to
 eliminate $\Delta x$ yields the equation for the Complementary Effect.
 By taking other projections the other entries of Table 3 may be calculated in 
 a similar fashion. It is interesting to note that the TD effect can be derived
 directly from the LFC effect by using the symmetry of the LT equations.
 Introducing the notation: $s \equiv ct$, the LT may be written as:
\begin{eqnarray}
 x' & = & \gamma (x-\beta s) \\
 s' & = & \gamma (s-\beta x)
 \end{eqnarray}
 These equations are invariant~\cite{x8} under the following transformations:
\begin{eqnarray} 
 T1~~~& : & x \leftrightarrow s,~~~ x' \leftrightarrow s' \\
 T2~~~& : & x \leftrightarrow x',~~~ s \leftrightarrow s',~~~ \beta
 \rightarrow -\beta     
\end{eqnarray}
Writing out the LFC entries in the first row of Table 3, replacing $t$, $t'$ by
$s/c$, $s'/c$ ; gives  
\[~~~\Delta x~~~~\Delta s = 0~~~~\Delta x = \frac{\Delta x'}{\gamma}~~~~
\Delta x' = -\frac{\Delta s'}{\beta}  \]
Applying $T1$ to each entry in this row results in:
\[~~~\Delta s~~~~\Delta x = 0~~~~\Delta s = \frac{\Delta s'}{\gamma}~~~~
\Delta s' = -\frac{\Delta x'}{\beta}  \]
Applying $T2$:
\[~~~\Delta s'~~~~\Delta x' = 0~~~~\Delta s' = \frac{\Delta s}{\gamma}~~~~
\Delta s = \frac{\Delta x}{\beta}  \]
Replacing $\Delta s$, $\Delta s'$ by $c\Delta t$, $c\Delta t'$
 yields the last row of Table 3 which describes
the TD effect. Similarly TC can be derived from SD (or vice versa) by successively
applying the transformations $T1$, $T2$.
\par The `Complementary Effects' listed in Table 3 have the following
 geometrical interpretations:
\begin{itemize}
\item LFC ($\Delta x' = -(c/\beta) \Delta t'$). This is the locus of all the
points in S' that are observed at the same time ($\Delta t = 0$) in S.
\item SD ($\Delta x = (c/\beta) \Delta t$). The locus of the moving object
as observed in S (see Fig1b).
\item TC ($\Delta x' = -c\beta \Delta t'$). The locus of the position
 of the local clock in S' observed at a fixed position ($\Delta x =0$) in S.
\item TD ($\Delta x = c\beta \Delta t$). The locus of the position
 of the moving local clock observed in S.
\end{itemize}
\par A remark on the `Observed Quantities' in Table 3. For the LFC, SD effects the
 observed
 quantity is a length interval in the frame S. The observed space distortion occurs
 because this length differs from the result of of a similar measurement made on
 the same object in its own rest frame. $\Delta x'$ is not directly measured
 at the time of observation of the LFC or SD. It is otherwise with the time
 measurements TD, TC. Here the time intervals indicated  {\it in their own rest frame}
 by a local moving clock (TD), or 
 different equivalent clocks at the same position in S (TC),  
 are supposed to be directly observed and compared
 with the time interval $\Delta t$ registered by an unmoving clock in the observer's rest frame.
 Thus the effect refers to two simultaneous observations by {\it the same observer}
 not to separate observations by {\it two different observers} as in the case 
 of the LFC and SD.

\par Einstein's first paper on Special Relativity~\cite{x1} showed, for the first time,
that the LFC and TD effects could be most simply understood in terms of the geometry of
space time, in contrast to the previous works of Fitzgerald, Larmor, Lorentz and Poincar\'{e}
 where
 dynamical and kinematical considerations were always mixed~\cite{x9}. However it can
 also be argued that Special Relativity has a dynamical aspect due to the changes in the
 electromagnetic field induced by the LT. Indeed, by calculations of the equilibrium
 positions of an array of point charges in both stationary and uniformly moving frames
 Sorensen has shown that the LFC may be derived from dynamical considerations~\cite{x10}.
 By considering several different `electromagnetic clocks' either at rest
 or in uniform motion, Jefimenko has demonstrated that the TD effect may also be
 dynamically derived~\cite{x11}. Similar considerations, emphasising the `dynamical'
 rather than the `kinematical' aspects of Special Relativity, have been presented in
 an article by Bell~\cite{x12}. Such calculations, based on the properties of 
 electromagnetic fields under the LT, demonstrate the consistency of Classical
 Electromagnetism with Special Relativity, but as pointed out by Bell~\cite{x12},
 in no way supersede the simpler geometrical derivations of the effects. It is not
 evident to the present author how similar `dynamical' derivations of the 
 new SD and TC effects could be performed.
 
 \par In conclusion the essential characteristics of the two `new' space-time 
 distortions discussed above are summarised :
 \begin{itemize} 
 \item \underline{Space Dilatation (SD):} If a luminous object lying along the Ox'
 axis, at rest in the frame S', is uniformly illuminated 
 for a short time  $\tau_L$ in this frame
 it will be
 observed from
 a frame S, in uniform motion relative to S' parallel to Ox' at the velocity -$\beta c$,
 in a direction perpendicular to the relative velocity, 
 as a narrow strip of width $c \tau_L/(\beta \gamma)$, perpendicular to the
 x-axis, moving with the velocity
 $c/\beta$ in the same direction as the object. The total distance swept out
 along the $x$-axis 
 by the strip during the time $\beta L/(c \sqrt{1-\beta^2})$, for which
 it is visible, is 
 $L/\sqrt{1-\beta^2}$ where $L$ is the length along Ox' of the object as observed in S'.
 Thus the apparent length of the object when viewed with a time resolution $\tau_R$
 much larger than $\beta L/(c \sqrt{1-\beta^2})$ is  $L/\sqrt{1-\beta^2}$.
 \item \underline{Time Contraction (TC): } The equivalent clocks in the moving frame 
 S', viewed at the same position in the stationary frame S, apparently run faster by a factor
 $1/\sqrt{1-\beta^2}$ relative to a clock at rest in S.
\end{itemize}
\begin{table}
\begin{center}
\begin{tabular}{|p{1.45in}|c|c|c|c|} \hline
 Name & Observed Quantity & Projection & Effect & Complementary Effect \\
\hline
Lorentz-Fitzgerald Contraction (LFC) & $\Delta x$ & $\Delta t = 0$   
& $\Delta x = \frac{1}{\gamma} \Delta x'$  & $\Delta x' = - \frac{c}{\beta} \Delta t'$ \\
\hline
\mbox{Space Dilatation} (SD) & $\Delta x$ & $\Delta t' = 0$ 
 & $\Delta x = \gamma \Delta x'$  & $\Delta x = \frac{c}{\beta} \Delta t$ \\
\hline 
\mbox{Time Contraction} (TC) & $\Delta t'$ & $\Delta x = 0$   
& $\Delta t' = \gamma \Delta t$  & $\Delta x' = - c \beta \Delta t'$ \\
\hline
\mbox{Time Dilatation} (TD) & $\Delta t'$ & $\Delta x' = 0$   
& $\Delta t' = \frac{1}{\gamma} \Delta t$  & $\Delta x =  c \beta \Delta t$ \\
\hline
\end{tabular}
\caption[]{ Different observed distortions of space-time in Special Relativity
(see text).  }      
\end{center}
\end{table}

 {\bf Acknowledgements} 
\par I thank G.Barbier and C.Laignel for their valuable help in the preparation
 of the figures, and an anonymous referee whose pertinent and constructive
 criticism has allowed me to much improve the presentation of Section 2.  
\pagebreak


\begin{thebibliography}{99}
\bibitem{x7}
If an observer in S' knows the distance $D$ to any of the clocks
 then the clock is synchronised relative to a local clock at the same position as the 
 observer, when it is observed to lag behind the latter by the time $D/c$ when viewed across
 free space.
\bibitem{x8}
 Actually the transformation $T2$ yields
 the inverse of the LT (5.7),(5.8). The inverse equations may then be solved to
 recover (5.7) and (5.8).
\end{thebibliography}
\end{document}